# Quantum Phase Transition in Pr$_2$CuO$_4$ to Collinear Spin State in Inclined Magnetic Field: A Neutron Diffraction Study.

V.P. Plakhty[1], S.V. Maleyev[1], S.V. Gavrilov,[1], F. Bourdarot[2,3], S. Pouget[2,3], and S.N. Barilo[4]

[1]*Petersburg Nuclear Physics Institute RAS, Gatchina, 188350 St.Petersburg, Russia*
[2]*CEA-Grenoble, DRFMC/SPSMS/MDN, 38042 Grenoble Cedex 9, France*
[3]*Institut Laue-Langevin, BP 156, 38042 Grenoble Cedex 9, France*
[4]*Institute for Solid State Physics and Semiconductors, Minsk, Belarus*

In the external field slightly inclined to the **x**- or **y**-axis of the frustrated tetragonal antiferromagnet Pr$_2$CuO$_4$, a transition is discovered from the phase with orthogonal antiferromagnetic spin subsystems along [1,0,0] and [0,1,0] to the phase with the collinear spins. This phase is shown to be due to the pseudodipolar interaction, and transforms into the spin-flop phase (**S**⊥**H**) asymptotically at very high field. The discovered phase transition holds at $T = 0$ and is a quantum one, with the transition field being the critical point and the angle between two subsystems being the order parameter.

Rare-earth cuprates $R_2$CuO$_4$ ($R$ = Pr, Nd, Sm, Eu, Gd), which originally had drown attention as parent compounds for the electron-doped high-$T_C$ superconductors, now are being extensively studied as two-dimensional quantum Heisenberg antiferromagnets. These materials have a tetragonal structure with the space group *I*4/*mmm*. The CuO$_2$ planes give a motif of the structure, with the Cu$^{2+}$ ions being coordinated by regular squares of oxygen neighbours [1]. A very important feature of this tetragonal body-centred structure is a shift of adjacent CuO$_2$ planes by [1/2,1/2,1/2]. Due to very strong in-plane superexchange interaction, $J$ = 124(3) meV [2], each Cu$^{2+}$ ion ($S$ = 1/2) has four nearest neighbors with the opposite spins. Therefore, the mean field, produced on each copper ion by the adjacent plains is cancelled. In the absence of exchange field, the interplanar spin orientation should be very sensitive to any weak interaction that violates the symmetry. Shender [3] has shown that in a similar situation of the bcc lattice the quantum zero-point spin fluctuations stabilize a collinear orientation of the spin subsystems, which has been confirmed experimentally [4].

For interpretation of the early neutron diffraction data a collinear model was used with the spins along [1,1,0] and with a propagation vector $\mathbf{k}_1 = (\pi/a)[1,1,0]$ in the same direction [5, 6]. Certainly there should be another domain with the spins along [1,−1,0] and with $\mathbf{k}_2 = (\pi/a)[1,−1,0]$. A different model with two vectors, $\mathbf{k}_1$ and $\mathbf{k}_2$ (2-**k** structure) that results in exactly the same intensities of the magnetic reflections has been also proposed [6]. In this model two antiferromagnetic subsystems are orthogonal, with the spins in positions (0,0,0) and (1/2,1/2,1/2) along [1,0,0] and [0,−1,0] directions, respectively, as shown in Fig. 1(a). Neutron diffraction experiments in the external magnetic field applied in the [1,1,0] direction [7-11] have definitely proved that the copper spin subsystems are orthogonal in spite of the quan-tum effect [3]. This structure is realized in Pr$_2$CuO$_4$ and in the phases I, III of Nd$_2$CuO$_4$ [7,12-14]. The opposite direction [0,1,0] of the (1/2,1/2,1/2) spins corresponds to another structure that exists in the cuprates of Sm [8], Eu [9,10,14], Nd (phase II) [12-14] and results in completely different intensities of the magnetic reflections.

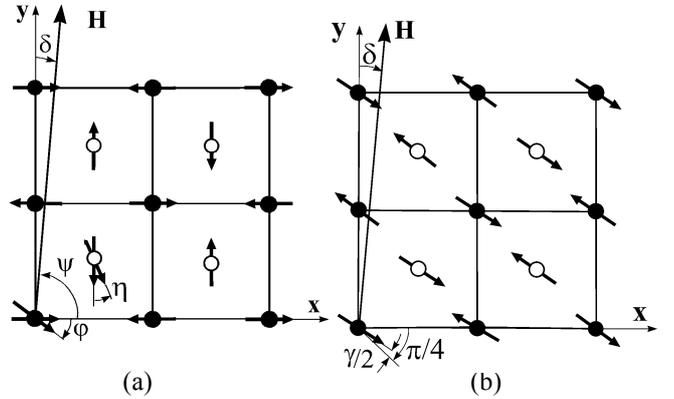

FIG. 1. (a) Spin structure of Pr$_2$CuO$_4$ in zero magnetic field. Two copper subsystems at $z = 0$ and $z = 1/2$ are described by the filled and open circles, respectively. The corresponding spin rotation by the angles $\varphi$ ($\eta$) in the field inclined to the **x** (**y**) axis by the angles $\psi$ ($\delta$) is also shown. (b) Spin structure of Pr$_2$CuO$_4$ in the collinear phase. Angle $\gamma$ is defined in the text.

These spin structures were explained by various models [12,15,16], but a relevant one seems to be based on the pseudodipolar interaction (PDI) postulated by Van Vleck in 1937. It has the same symmetry as the magnetic dipolar one, but may be much stronger. It has the form:

$$V_{\mathrm{PD}} = \frac{1}{2}\sum_{nn'} V(\mathbf{R}_{nn'})(\mathbf{S}_{n'}\hat{\mathbf{R}}_{nn'})(\mathbf{S}_n\hat{\mathbf{R}}_{nn'}), \qquad (1)$$

where $n$, $n'$ label the lattice point positions, and $V(R)$ is steeper than $R^{-3}$ for the dipolar interaction at $R \to \infty$. Moria has shown that the PDI is a result of the superexchange in presence of the spin-orbit coupling [17]. For cuprates it was extensively studied theoretically. (See [18] and references therein.) However in these studies the intraplane PDI was the only evaluated. The role of the PDI between adjacent $CuO_2$ planes in the $R_2CuO_4$ materials in connection to their non-collinear spin structure was recognized in Refs. [19,20]. In particular the spin-reorientation transitions in $Nd_2CuO_4$ were explained by assuming two superexchange pathways with one and two $Nd^{3+}$ Kramers ions between the $Cu^{2+}$ ions in the adjacent planes. The intraplane PDI gives rise to the square anisotropy in the $CuO_2$ planes that is responsible for the in-plane spin-wave gap $\Delta_0$ [21].

The spin-wave spectrum in presence of both intraplane and interplane PDI's has been evaluated, and the data of the inelastic neutron scattering experiments for $Pr_2CuO_4$ have been presented in Ref. [20]. The parameters $P$ and $Q$ of the nearest-neighbor PDI's in a plane and between the planes have been found, respectively, from the $\Delta_0$ and the splitting $\Omega_{opt}$ of the in-plane spin waves on the acoustic and optic branches due to the PDI. On the other hand these parameters can be obtained from the field dependence of the spin structure. If the field is applied along the [1,1,0] direction, the spin-flop transition at low temperature is sharp and looks like a second order one [11,22] at the critical field $H_C[1,1,0]$. In the case when magnetic field is applied along [1,0,0], or [0,1,0] this transition is expected [20] to be of the first order at $H_C[1,0,0]$. From these critical field values the parameters $P$, $Q$ and $\Delta_0$ can be determined using Eqs. (2, 8, 40, 93) of Ref. [20]. If the angle between the field and the [0,1,0] direction, $\delta \ll 1$, the deviations $\varphi$ and $\eta$ of two subsystems from their zero-field directions [1,0,0] and [0,–1,0] is given in the first approximation for small field as

$$\varphi = -\eta \approx K^2 \delta, \qquad (2)$$

where $K = g\mu_B H / \Delta_0$ [23].

The objective of this experiment was a determination of the PDI-parameters from the $H_C[1,0,0]$, $H_C[1,1,0]$ and from the dependence (2) to check the consistency with the values obtained by inelastic neutron scattering [20]. However, when measuring the field dependence of $\varphi$ and $\eta$ at $\delta = 9.5°$, we have observed a novel transition to the collinear spin structure. Therefore, here we have concentrated on this point, leaving the other results to be presented elsewhere.

A $Pr_2CuO_4$ single crystal (I) with mosaicity of about 0.2° that we have studied in the present experiment was grown in air from the melt in crucible, and the growing conditions were similar to those for the crystal II used in the spin-wave study [20]. Our crystal has a plate-like shape with the dimensions of about $10 \times 5 \times 1$ mm$^3$. The lattice parameters at $T = 300$ K are $a = 3.945$ Å, $c = 12.16$ Å. From the integrated intensity temperature dependence of magnetic reflection (1/2,1/2,–1) the Néel point has been found as $T_N = 250(3)$ K, i.e., practically the same as $T_N = 247$ K for the crystal II. This is important for comparison the results, since the crystal properties are very sensitive to the growing procedure, apparently due to oxygen non-stoichiometry.

The crystal had been mounted in a cryomagnet with vertical field up to 10 T installed on D15 diffractometer with a lifting detector at the reactor of Institut Laue-Langevin. The axis [0,1,0] was vertical with a precision of 0.15°. Field dependence of the integrated intensity of the magnetic reflection (–1/2,1/2,–1) was measured at $T = 18$ K, with the field increasing from zero to 8 T and then decreasing back to zero. Before the intensity measurement the field was stabilized during a half of an hour. In Fig. 2, the transition looks like the first order as predicted in Ref. [20] where the high-field phase was assumed to be the spin-flop one. Non-zero intensity in the flopped phase can be due to an admixture of different domains. The field value, corresponding to the average intensity, was taken for the critical field $H_C[1,0,0] = 5.42(1)$ T with the hysteresis, $\Delta H = 0.08(1)$ T.

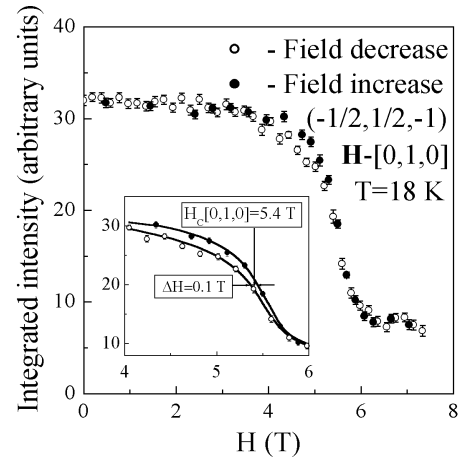

FIG. 2. Spin-reorientation transition in the external field along the [0,1,0] direction. The hysteresis is shown in the inset.

Unfortunately we were not able to repeat the measurement [22] of $H_C[1,1,0]$ with a new crystal I. From the dependence (2) in small field inclined to the y-axis by $\delta = 9.5°$ (Fig. 1), we have estimated $H_C[1,1,0] \approx 1.7$ T at $T = 18$ K, which is in evident contradiction with $H_C[1,1,0] = 3.11(3)$ T that follows from the spin-wave gap $\Delta_0 = 0.36(3)$ meV [20].

…All further measurements of the intensities $I_{(-1/2,1/2,-1)}$ and $I_{(1/2,12,-1)}$ have been performed in the field inclined by 9.5°. These intensities can be calculated according to [24] as

$$I_{(h,k,l)} = \left|\mathbf{F}_{(h,k,l)}\right|^2 - (\hat{\mathbf{e}}_{(h,k,l)} \mathbf{F}_{(h,k,l)})^2, \qquad (3)$$

where $\mathbf{F}_{(h,k,l)}$ – is the magnetic structure amplitude, and $\mathbf{e}_{(h,k,l)}$ – is the unit scattering vector in the direction of the momentum transfer. The intensity is given in arbitrary units.

We calculate the intensity for the unit cell $2\mathbf{a}\times2\mathbf{a}\times\mathbf{c}$ shown in projection along [0,0,1] in Fig.1, but use the half-integer indices $(-1/2,1/2,-1)$ and $(1/2,1/2,-1)$ to label the reflections. In the inclined field, the spins of the first subsystem with $z = 0$ are turned by an angle $\varphi$ while the spins of the second one with $z = 1/2$ by an angle $\eta$ from their original directions along [1,0,0] and [0, −1,0], respectively. From (3) it follows that $I_{(\pm1/2,1/2,-1)} = 32$ at $H = 0$ when $\varphi = \eta = 0$. We shall introduce the angles $\alpha$ and $\gamma$ as $\varphi = \alpha + \gamma/2$ and $\eta = -\alpha + \gamma/2$. Using Eq. (3) these angles can be expressed through the intensities $I_{(\pm1/2,1/2,-1)}$ as

$$\cos\gamma = 1 - [2 + (2a/c)^2]\cdot[1 - (I_{(-1/2,1/2,-1)} + I_{(1/2,1/2,-1)})/64], \quad (4)$$
$$\sin 2\alpha = (I_{(-1/2,1/2,-1)} - I_{(1/2,1/2,-1)})/(I_{(-1/2,1/2,-1)} + I_{(1/2,1/2,-1)}). \quad (5)$$

It follows from Eq. (4) that in zero field ($\varphi = \eta = 0$) both reflections have equal intensities $I_{(1/2,1/2,-1)} = I_{(-1/2,1/2,-1)} = 32$, while in the spin-flop phase, when both spin subsystems are perpendicular to $\mathbf{H}$, $I_{(-1/2,1/2,-1)} = 0$ and $I_{(1/2,1/2,-1)} = 45.7$ at $\delta = 9.5°$. The field dependence of these intensities normalized to 32 at $H = 0$ is displayed in Fig. 3. Two features are seen:
1 − there is a new transition, apparently of the second order, at $H_C \approx 2.9$ T;
2 − the intensity $I_{(1/2,1/2,-1)} \approx 62$ above this transition is considerably higher than it should be in the spin-flop phase and changes only a little up to $H = 9.3$ T, well above the $H_C[100]$. Together with the $I_{(-1/2,1/2,-1)} \approx 0$ it corresponds to $\eta \approx -\varphi \approx 45°$.

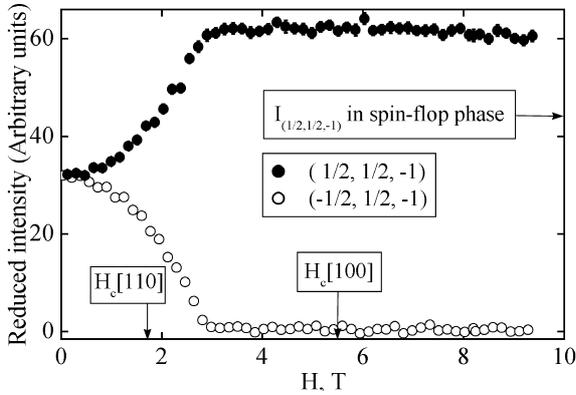

FIG. 3. Field dependence of the intensity of magnetic Bragg reflections (1/2,1/2,−1) and (−1/2,1/2,−1) normalized to 32 at $H = 0$. The intensity $I_{(1/2,1/2,-1)}$ in the spin-flop phase is indicated on the right. The critical fields of the spin-flop transition $H_C[110]$ (second order) and of the $H_C[100]$ (first order) are also shown.

Using Eqs. (4, 5) the field dependences of $\alpha$ and $\gamma$ shown in Fig. 4 have been calculated. From Eq. (5) one can obtain two sets of $\alpha$ with $2\alpha > -\pi/2$ and $2\alpha < -\pi/2$. The difference between corresponding points in Fig. 4 gives an uncertainty of $\alpha$ at $\alpha \approx -\pi/4$. It is seen that $\alpha = -\pi/4$ above $H_C$ in all the range studied up to $H = 9.3$ T. This means that in the new phase the spins subsystems are collinear, since their angles with the $\mathbf{x}$-axis are equal ($\varphi_1 = \varphi = -\pi/4 + \gamma/2$, and $\varphi_2 = -\pi/2 + \eta = -\pi/4 + \gamma/2$) as shown in Fig. 1(b).

This on the first sight unexpected result may be explained,

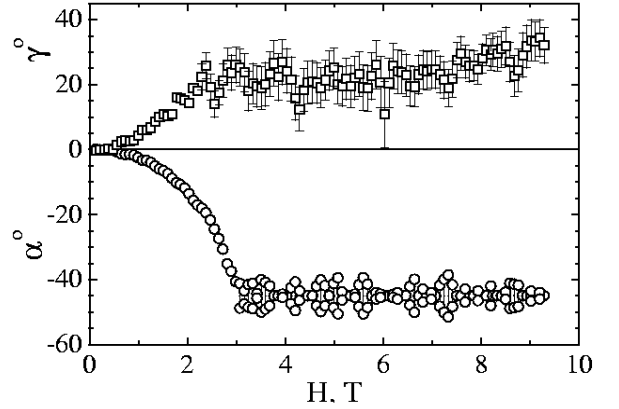

FIG. 4. Field dependences of the angles $\alpha$ (circles) and $\gamma$ (squares) defined in the text.

at least qualitatively, using classical energy of the easy-plane frustrated tetragonal antiferromagnet in magnetic field [20]. The contributions to this energy important for our consideration (the square anisotropy of two sets of the $CuO_2$ planes, the interplane pseudodipolar energy and the magnetic energy) are represented, respectively, by the first, second and third terms in the expression:

$$E = E_0\{\sin^2 2\varphi + \sin^2 2\eta)/4 - G\cos(\varphi + \eta) - K^2[\sin^2(\psi - \varphi) + \cos^2(\psi - \eta)]\}, \quad (6)$$

where $G = (\Omega_{opt}/\Delta_0)^2$. The prefactor is expressed through the total number $N$ of the unit cells, the spin $S$ and the in-plane exchange parameter $J$ as $E_0 = \Delta_0^2 N/(16SJ)$. Eq. (6) coincides with the first five terms in Eq. (85) of Ref. [20]. (All the other terms of this equation are wrong, as the square anisotropy has not been correctly taken into account.)

Stability of the spin structure is determined by conditions: $\partial E/\partial\varphi = 0$, $\partial E/\partial\eta = 0$ and $(\partial^2 E/\partial\varphi^2)(\partial^2 E/\partial\eta^2) - (\partial^2 E/\partial\varphi\partial\eta)^2 \geq 0$. The last inequality coincides with the condition of the spin-wave stability, which can be shown using the results of Appendix B in Ref. [20]. These conditions applied to Eq. (6) give the following expressions:

$$[\sin 2\alpha \cdot \cos 2\gamma + K^2\sin(2\delta + \gamma)]\cos 2\alpha = 0, \quad (7)$$

$$\cos 4\alpha \cdot \sin 2\gamma + 2K^2\sin(2\delta + \gamma)\sin 2\alpha + 2G\sin\gamma = 0, \quad (8)$$

$$[\cos(4\alpha + 2\gamma) + K^2\cos(2\delta + 2\alpha + \gamma)]$$
$$\times [\cos(4\alpha - 2\gamma) - K^2\cos(2\delta - 2\alpha + \gamma)]$$
$$+ (G/2)[\cos(4\alpha + 2\gamma) + \cos(4\alpha - 2\gamma)]\cos\gamma$$
$$+ (K^2/4)[\cos(2\delta + 2\alpha + \gamma) - \cos(2\delta - 2\alpha + \gamma)] \geq 0. \quad (9)$$

It follows from Eqs. (7-9) that at $\delta \neq 0$ and in a weak field

$$K_C^2 \cos(2\delta + \gamma_C) = (G - \cos\gamma_C)\sin\gamma_C, \quad (10)$$
$$K_C^2 \cos(2\delta + \gamma_C) = \cos 2\gamma_C. \quad (11)$$

For $H > H_C$ the angle $\gamma$ can be obtained from (10), if one omits the subscript "C", and it is seen that the spin-flop state with $\mathbf{S} \perp \mathbf{H}$ is gained at $H \to \infty$.

Eqs. (7-9) hold also in the case $\delta = 0$ ($\mathbf{H} \parallel [0,1,0]$), for which a transition to the spin-flop state was assumed [20] at $g\mu_B H_C[100] = \Delta_0(1 + G)^{1/4}$. Actually, analysis of (7-9) shows that this is a first order transition from the state with $\alpha = \gamma = 0$ to the collinear phase with $\alpha = -\pi/4$ again, but with $\gamma$ determined by Eq. (11) with $H_C[100]$ substituted by $H$. In the new phase $\alpha$ remains constant while $\gamma$ increases as $\tan\gamma = K^2/(G - \cos\gamma)$. In our case $G \approx 60$, $\tan\gamma \approx (K/\Omega_{opt})^2$, and the spin-flop state is gained at $H \to \infty$ as before. In both cases the transition to the collinear state $\alpha = -\pi/4$ is driven by the interplane pseudodipolar interaction that is characterized for $Pr_2CuO_4$ by the large $G$.

Coming back to the second order transition at $H_C$, we should point out that expression (9) is positive on both sides of the critical point and becomes zero at $H = H_C$. This means that the in-plane acoustic spin-wave branch becomes gapless at the critical point. Here we deal with a quantum phase transition as it holds at $T = 0$, and $H_C(\delta)$ is a quantum critical point [25]. The state at $H > H_C$ may be considered as a disordered one, and the order parameter is the angle between the spins at $z = 0$ and $z = 1/2$ equal to $\pi/2 + 2\alpha$ (Fig. 1a). At $H = H_C$ interaction between the gapless spin waves becomes important, $H_C$ is renormalized, and the mean-field energy (4) needs a quantum correction, as it is seen from comparison of two values of $H_C[1,1,0]$. Hence the theory outlined above explains the experimental data only qualitatively.

In conclusion, we have observed by means of neutron diffraction a novel magnetic transition in the frustrated layered cuprate $Pr_2CuO_4$. The spin structure with orthogonal antiferromagnetic subsystems transforms into a collinear one by a transition continues in the external field slightly inclined to the axis of one subsystem. Analysis of the classical energy of the easy-plane frustrated tetragonal antiferromagnet in magnetic field [20] shows that this transition is driven by the pseudodipolar interaction. It is the first order one when the field is applied exactly in the direction of one spin subsystem. In both cases the spin-flop state is gained only at $H \to \infty$. The transition is a quantum one as it holds at $T = 0$, with $H_C(\delta)$ being the quantum critical point and the angle between subsystems in the ordered phase $H < H_C$ being the order parameter. The interaction between the gapless spin waves at the critical point strongly renormalizes the value of $H_C$, and our theory needs quantum corrections.

We kindly acknowledge partial support from the Russian Foundations for Fundamental Researches (Projects Nos. 99-02-17273, 00-02-16873, 00-15-96814, 02-02-16981), Russian State Programs for Statistical Physics and Quantum Macrophysics, Ministry of Science (theme N° 4).